\documentclass[journal,comsoc]{IEEEtran}

\ifCLASSINFOpdf

\else

\fi

\usepackage{mathrsfs}
\usepackage{amsmath}
\usepackage{amsfonts}
\usepackage{amsthm}
\usepackage{amssymb}
\usepackage{graphicx}
\usepackage{subfigure}
\usepackage{indentfirst}
\usepackage{array}
\usepackage{epstopdf}
\usepackage{cite}
\usepackage{enumerate}
\usepackage{bm}
\usepackage{dsfont}
\usepackage[linesnumbered,ruled,vlined]{algorithm2e}
\usepackage{multirow}
\usepackage{verbatim}
\usepackage{stfloats}
\usepackage{color}

\SetKwComment{Comment}{//}{}
\SetKwBlock{Begin}{Function}{end}

\begin{document}

\title{\huge{Towards Semantic Communications: A Paradigm Shift}}

\author{Kai Niu, \IEEEmembership{Member, IEEE},
Jincheng Dai, \IEEEmembership{Member, IEEE},
Shengshi Yao, \IEEEmembership{Student Member, IEEE},
Sixian Wang, \IEEEmembership{Student Member, IEEE},
Zhongwei Si, \IEEEmembership{Member, IEEE},
Xiaoqi Qin, \IEEEmembership{Member, IEEE},
Ping Zhang, \IEEEmembership{Fellow, IEEE}



\thanks{The authors are Beijing University of Posts and Telecommunications.}

\vspace{0em}

}

\maketitle

\begin{abstract}
The last seventy years have witnessed the transition of communication from Shannon's theoretical concept to current high-efficient practical systems. Classical communication systems address the capability-deficiency issue mainly by module-stacking and technique-densification with ever-increasing complexity. In such a traditional viewpoint, classical source coding only uses explicit probabilistic models to compress data, regardless of the meaning of transmitted source messages. Also, channel coded transmission does not identify the source content. In this sense, state-of-the-art communication systems work merely at the technical level as summarized by Weaver. Unlike the traditional system design philosophy, this article proposes a new route to boost the system capabilities towards intelligence-endogenous and primitive-concise communications. The communication paradigm upgrades to the semantic level, which is radically different since all the key techniques imply the use of meanings of transmitted data, thus deeply changing the design of the communication system. This paradigm shifting unveils a promising direction due to its ability to offer an identical quality of service with much lower data transmission requirement. Different from other similar works, this article constitutes a brief tutorial on the framework of semantic communications, its gain analyzed from the information theory perspective, a method to calculate the semantic compression bound, and an exemplary use case of semantic communications.
\end{abstract}

\IEEEpeerreviewmaketitle

\section{Introduction}\label{section_introduction}

In the past decades, modern communications have evolved from the Shannon's theoretical concepts to practical systems enabling the Internet of Everything (IoE). With the rapid development of mobile communications, there gradually forms a thinking set: the system bottleneck is in the channel capacity, and increasing capacity can address most problems. In this way, the wireless network evolution was primarily driven by the need for higher data rates, essentially via the use of broadband (multi-carriers or OFDM) combined with high-performance channel coding, high-order modulation, large-scale MIMO, etc. One can argue today's outward evolution route has gained significant momentum from 1G to 5G. However, current communication systems are still working merely at \emph{LEVEL-A} -- technical level -- guided by the Shannon information theory \cite{thomas}, where all problems are formulated with explicit probabilistic models.

Future communications will embrace artificial intelligence (AI) as one of key ingredients, and AI-empowered services will evolve into two setups: the real world and the virtual world. The former one is compatible with current communication scenes and infrastructures, while the latter one additionally deals with novel virtual-world requirements. To bridge the real and the virtual world, we embrace a new element -- Genie as an intelligent agent, in addition to the existing three physical elements, i.e., Humans, Machines, and Things. The Genie can aggregate and extract valuable information from physical communication objects and enable more efficient purpose-oriented interaction among communication objects. Thus, built upon the integration of AI and next-generation networking technologies, the future network interactions will evolve to the ``AI + Internet of Services (AIoS)'' era, which will be concise and efficient by using the intelligent-enough Genie to identify the task-related information and judiciously exchange the vital information only. As a new adjective, we use ``intellicise'' to stand for intelligence-endogenous and primitive-concise. Such an intellicise communication paradigm emphasizes understanding the communication goals and contents. To this end, the communication system comes to the point of upgrading to the \emph{LEVEL-B} -- semantic level -- as predicted by Weaver in 1949 \cite{weaver1953recent}.

Recently, semantic communications are emerging as a new paradigm driving the arrival of AIoS \cite{qin2021semantic, zhang2021toward,deepsc,shi2021semantic, luo2022semantic, DJSCC,NTSCC}. As for the physical layer, recent advances in computer vision and natural language processing enable semantics-oriented communications, which well boost the transmission quality and efficiency of various source modalities. As for the media access control, link, and network protocol layers, the redundancy of the layered protocols has been reduced, benefiting from the semantic-filtering mechanism. As for the application layer, semantics-based user intent identification has been explored to be applied to network management and configurations. Although we are witnessing great success in implementing semantics-aware technologies in each layer of communication networks, a systematic framework of semantic communications is still missing.

In this article, we elaborate on the insight of integrating semantic information processing techniques with wireless communications, present an overview of the general framework of semantic communications, analyze its gain from the information theory perspective, calculate the semantic compression bound with regard to Kolmogorov complexity, and demonstrate an exemplary use case of semantic communications.

\section{Challenges and Motivation}\label{section_challenges}

\subsection{Challenges in Classical Communications}

As Shannon and Weaver stated in \cite{weaver1953recent}, the fundamental problem of communication at \emph{LEVEL-A} -- technical level -- is that of reproducing the symbols exactly sent by the transmitter. Guided by this, the main effort in the long-term evolution of communication has always been focusing on the effectiveness and reliability of transmission. The whole system has been explicitly divided as source part and channel part.

In order to indicate the characteristic of source, entropy, and rate-distortion function are introduced to describe the lower bound of compressed symbol length such that source symbols can be exactly recovered or reconstructed within some distortion, corresponding to the lossless source coding theorem and the lossy source coding theorem of the Shannon theory. Various source coding methods or standards have been proposed for lossless source coding, such as Huffman coding, arithmetic coding, etc., and lossy source coding, such as pulse code modulation (PCM), code-excited linear prediction (CELP), MPEG-1, etc., for audio signals, and JPEG, BPG, etc., for image signals, and MPEG-4, H.264, H.265, etc., for video signals. After long-term evolution, the compression rates of lossy source coding have been quite closely approaching their rate-distortion functions.

On the other hand, the Shannon limit describes the lowest signal-to-noise ratio (SNR) required to error-free transmit messages at the given rate. As the most important technique approaching the Shannon limit, channel coding has been developing from primitive Hamming codes, convolutional codes, to advanced Turbo codes, low-density parity-check (LDPC) codes, polar codes, etc. Meanwhile, a lot of techniques have been proposed to increase spectrum efficiency, e.g., high-order modulation, MIMO, etc. Combining these with advanced channel codes can achieve a high achievable transmission rate. According to Shannon's noisy channel coding theorem, in order to fulfill the high-bandwidth utilization, any further increase in the data rate requires a significant augmentation in the received signal power unless the bandwidth is extended in proportion to the incremental data rate. Guided by this, mobile communication systems are evolving under the outward-expansion mode in the following ways.
\begin{itemize}
  \item Extend the bandwidth: this approach has been adopted from 2G to 5G, increasing the bandwidth from 5MHz to 100MHz within the sub-6G band, and we are now trying the Millimeter wave and even the Terahertz band.

  \item Increase the signal-to-noise ratio (SNR): this is achieved by reducing the cell coverage radius which forms the picocell structure, using beamforming, interference-suppress techniques, etc.

  \item Add the number of parallel channels: this is realized by using large-scale MIMO, OFDM, advanced multiple access techniques, etc.
\end{itemize}

These typical technologies have achieved tremendous success in today's communication systems, whereas they also lead to severe high-frequency coverage costs, complicated signal processing, high energy consumption, etc. The traditional path to approach the channel capacity or increase spectrum efficiency is unsustainable and will hit a plateau early or late. Furthermore, guided by the Shannon separation principle \cite{thomas}, almost all current source and channel modules are designed based on the separation idea. This paradigm allows for diverse sources to share the same digital media, i.e., classical communication systems only consider conveying the source coded bit sequence naively and reliably from transmitter to receiver. For finite-blocklength communications, however, the separation principle cannot be fulfilled \cite{thomas}, especially for massive short to moderate packet transmissions where the asymptotic limits can hardly be achieved. In addition, the channel transmission does not consider the content meaning inside the source. Just as commented by George Bernard Shaw ``\emph{The biggest problem in communication is the illusion that it has taken place}'', naively increasing channel capacity cannot address all communication problems, especially in future intelligent eras.

\subsection{Going Intelligent and Concise: Semantic Communications}

The aforementioned discussions illustrate that traditional communication systems pursue an accurate bit-level transmission over a noisy communication channel. Shannon theory provides a fundamental theoretical limit on the rate of reliable communications, where the processing unit is bit. This approach has successfully served content delivery oriented wireless networks for decades. Nevertheless, the current communication paradigm works merely at the technical level, ignoring the meaning the bits convey or how they would be used. It comes to its shifting time to \emph{LEVEL-B} -- semantic level -- so as to transfer from the traditional outward-expansion developing mode to a new inward-discovery developing mode, as predicted by Weaver in \cite{weaver1953recent}.

\begin{figure}[htbp]
\setlength{\abovecaptionskip}{0.cm}
\setlength{\belowcaptionskip}{-0.cm}
 \centering{\includegraphics[scale=0.65]{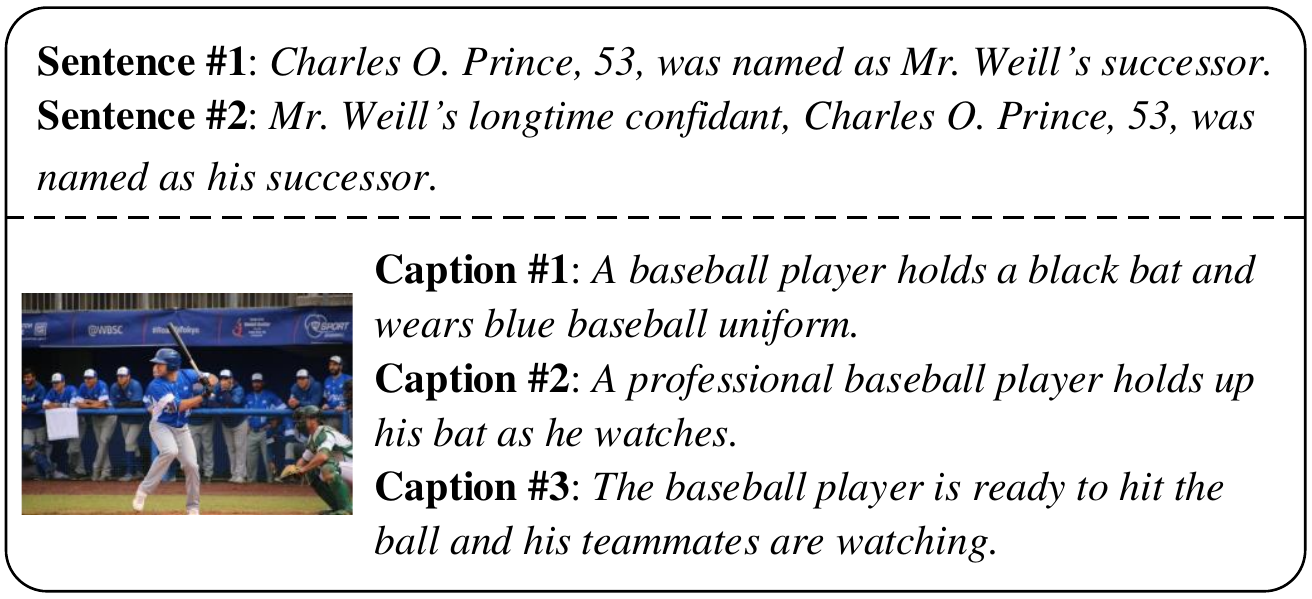}}
 \caption{Examples of expressions with similar semantic information.}\label{Fig1}
 \vspace{0em}
\end{figure}

\begin{figure*}[t]
\setlength{\abovecaptionskip}{0.cm}
\setlength{\belowcaptionskip}{-0.cm}
  \centering{\includegraphics[scale=0.4]{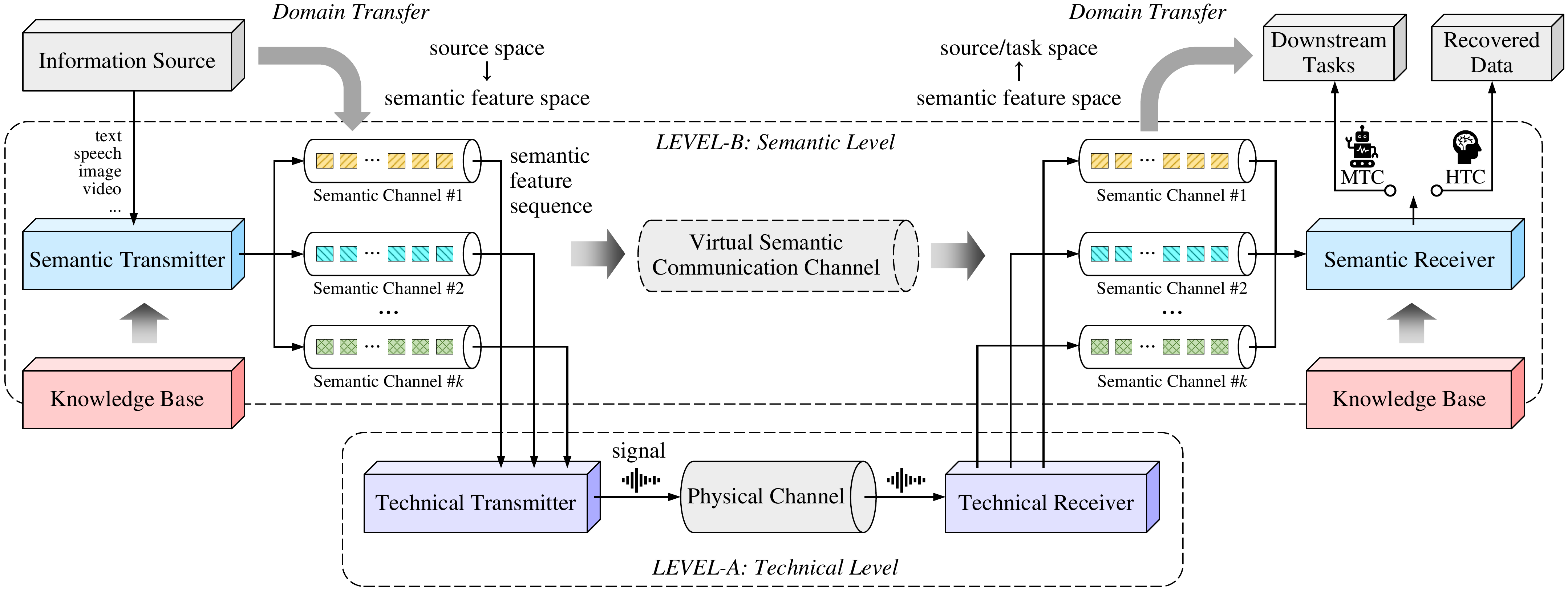}}
  \caption{Block diagram of semantic communication system for both human-type communications (HTC) and machine-type communications (MTC).}\label{Fig2}
  \vspace{0em}
\end{figure*}

At \emph{LEVEL-B}, similar semantic meaning could be presented with various expressions. Fig. \ref{Fig1} presents two examples. The sentence pair on the top constitutes paraphrases. They differ in the technical level but overlap in the semantic one. The bottom example lists a few descriptions of the image, which exist common ground while reserving differences. It follows that the semantics of the message can be of diversity and hierarchy. The similarity of the semantic meaning for the transmitter and receiver is what matters in the \emph{LEVEL-B} communication system. However, the problems at \emph{LEVEL-B} still require in-depth research.

In this new semantic paradigm, source content takes part in the communication system, interacting directly and seamlessly with other channel transmission modules also with intelligent functions. The ultimate goal is realizing the intellicise communication system. Recent advances in AI technologies and their applications have boosted the interest and the potentials of this communication paradigm shifting by processing and encoding semantics-related information, thus only the necessary information relevant to the specific task at the receiver will be transmitted. The higher-level semantic correlation inside source data could be captured instead of pure use of the statistical property. Apart from the source, wireless channel state information contains semantic information. The transmitter can select suitable modes to transmit the source semantic information. This methodology shifting in comprehensive understanding of source data and channel state finally leads to the change of transmission paradigm and shapes the semantic communication system.

\section{Semantic Communication System}

By a semantic communication system we mean a system of the type indicated schematically in Fig. \ref{Fig2}. It is a hierarchical structure consisting of both the technical level and the semantic level. Specifically, this system includes the following parts.

\begin{enumerate}[1.]
  \item An \emph{information source} that produces data to be transmitted to the destination. The information is implied by various modalities, e.g., text, audio, image, video, and their combinations. Data of different modalities are characterized by different levels of information density and redundancy.

  \item A local \emph{knowledge base} that provides additional guidance for the semantic transceiver. This background knowledge base exists in various formats: (a) knowledge graph; (b) database; (c) trained parametric or non-parametric models. Besides, it also provides \emph{a priori} knowledge about the channel state information for both transmitter and receiver. Such knowledge is used to facilitate the semantic feature extraction in the semantic transmitter, and it can be different for various transmission tasks.

  \item A \emph{semantic transmitter} that operates on the source data to extract the semantics-related features, which can be categorized into multiple semantic channels, and each one corresponds to one type of semantic feature of source data. As shown in Fig. \ref{Fig2}, for speech source, it can be decoupled as timbre, tone, content, etc. For image/video sources, using the semantic segmentation technique, the features of multiple semantic objects are extracted. In addition, each semantic feature corresponds to different semantic importance for specific tasks at the receiver, including data recovering or downstream intelligent tasks, such as language translation, object detection.

  \item A \emph{technical transmitter} that operates on the semantic feature sequence of each semantic channel in some way to produce signal suitable for transmission over the physical channel. It can utilize the traditional separation-based source and channel coding combined with modulation to produce digital signals. It can also directly generate the analog signals by using the emerging joint source-channel coding with deep neural networks.

  \item The \emph{physical channel} is merely the medium used to transmit the signal from transmitter to receiver.

  \item The \emph{technical receiver} ordinarily performs the inverse operation of that done by the technical transmitter, reconstructing the message from the signal.

  \item The \emph{semantic receiver} performs the inverse operation of that done by the semantic transmitter, exploiting the semantics fusion to reconstruct source data or directly execute downstream intelligent tasks.
\end{enumerate}

\section{From Shannon Information Theory to Semantic Information Theory}

Since the research of semantic communications is still at its infancy stage, there has been no consistent theory about semantic communications yet. Nevertheless, we can still generalize the Shannon information theory to conceptually understand the core idea of semantic communications.

Following the separation principle of Shannon information theory, most classical communication systems separate source processing and channel processing. Based on probabilistic modeling, the source compression stems from the concept of \emph{typical sequence} \cite{thomas}. It divides the set of all source sequences $S$ into two sets, the typical set and non-typical set. The sample entropy in the typical set is close to the true entropy, and the non-typical set contains the other sequences. Most of our attention will be on the typical sequences. Any property that is proved for the typical sequences will then be true with high probability and will determine the average behavior of a large sample. Therefore, we order all elements in the typical sequence set according to some order. Then we can represent each sequence of the typical set by giving the index of the sequence in the set. That forms the principle of classical source coding, after attaching redundancy bits in channel coding. Each source and channel coded sequence $X$ corresponds to a typical sequence in the source space $S$.

\begin{figure}[t]
	\setlength{\abovecaptionskip}{0.cm}
	\setlength{\belowcaptionskip}{-0.cm}
 \begin{center}
 \subfigure[Classical coded transmission]{
   \includegraphics[scale=0.59]{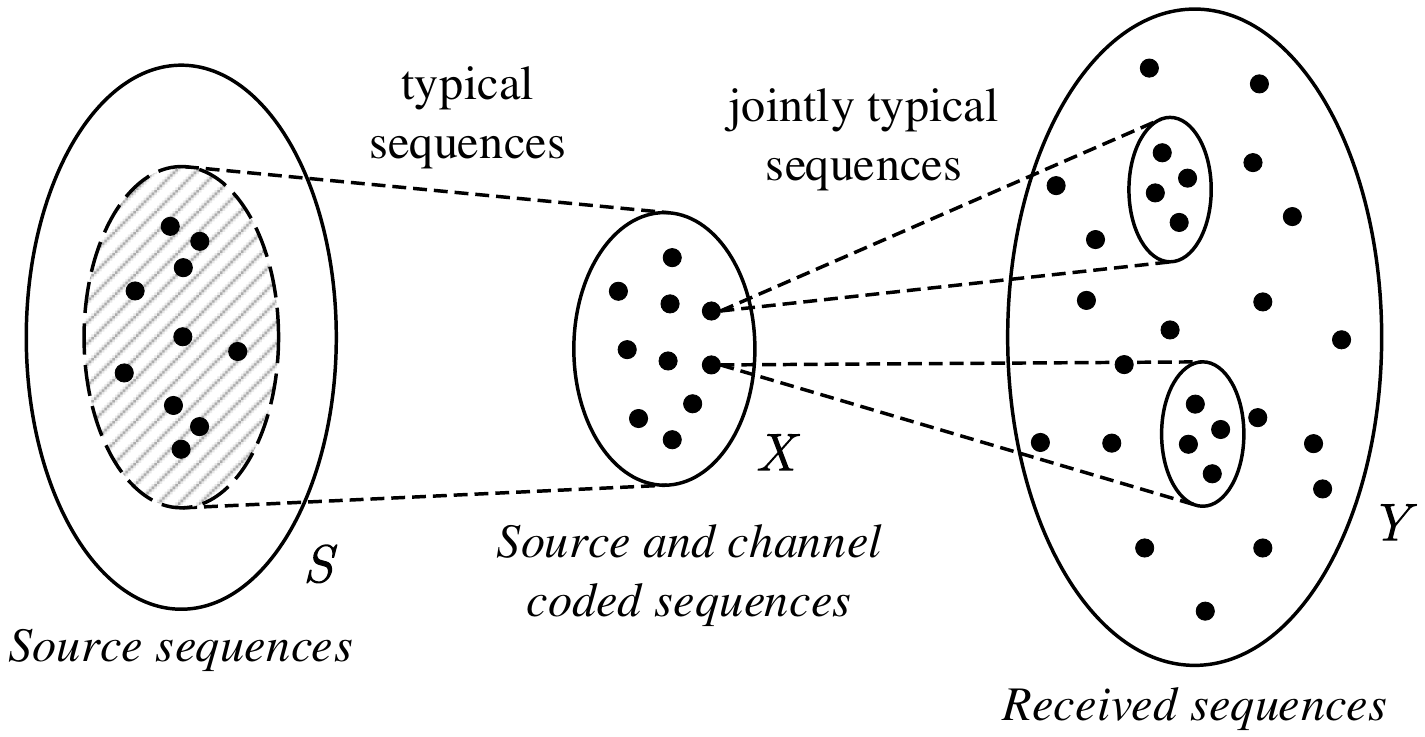}
 }
    \subfigure[Semantic coded transmission]{
 \includegraphics[scale=0.59]{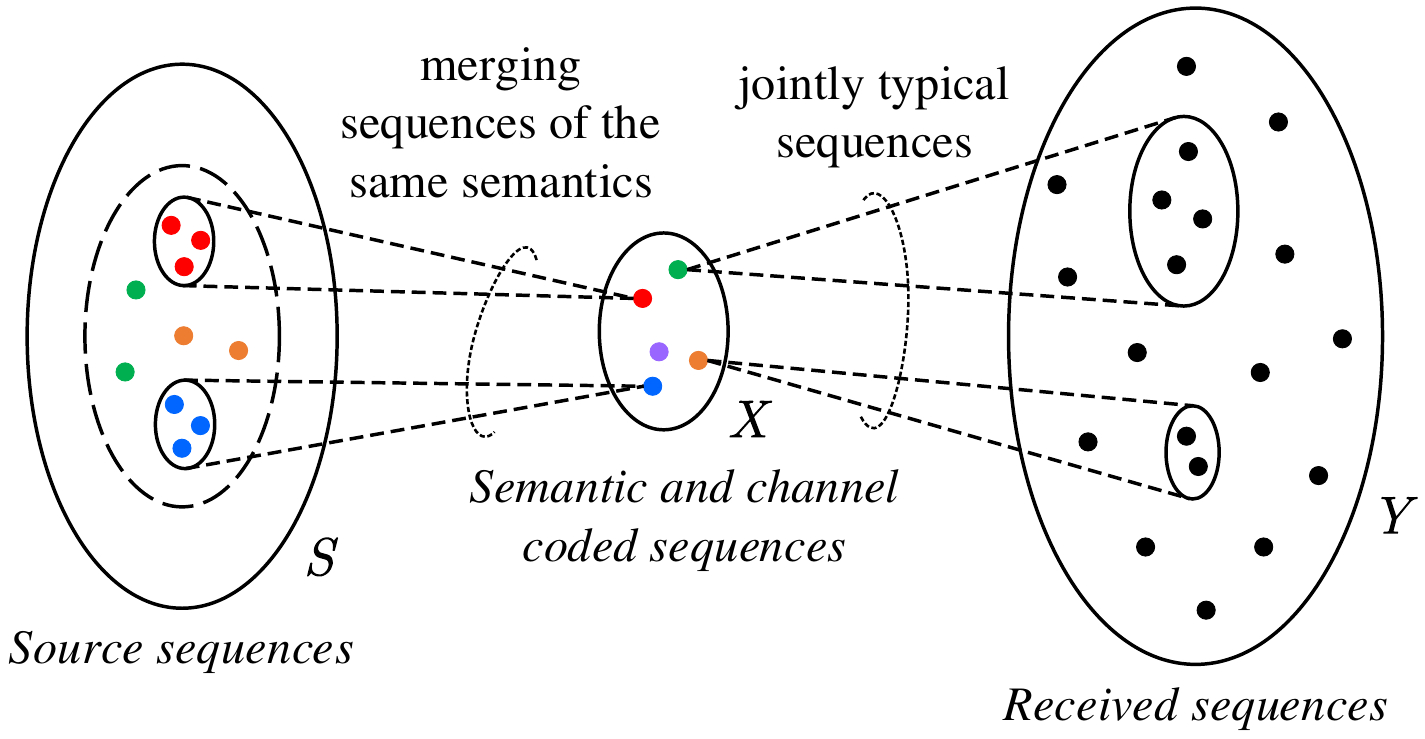}
 }
 \caption{A comparison between classical coded transmission and semantic coded transmission from typical sequence perspective.}
 \label{Fig3}
 \end{center}
\end{figure}

In terms of channel transmission, for each (typical) channel input sequence, the number of possible $Y$ sequences depends on the conditional entropy conditioned on $X$, all of them equally likely. Looking at Fig. \ref{Fig3}(a) in terms of typical sets, we see that associated with every $X$ is a typical ``fan'' of $Y$ sequences that are jointly typical with the given $X$ as shown in Fig. \ref{Fig3}(a). We wish to ensure that no two $X$ sequences produce the same $Y$ output sequence. Otherwise, we will not be able to decide which $X$ sequence was sent. The total number of possible (typical) $Y$ sequences depends on the entropy of $Y$. This set has to be divided into subsets corresponding to the different input $X$ sequences. Therefore, the number of distinguishable sequences we can send at most depends on the mutual information between the channel input and output \cite{thomas}.

As depicted in Fig. \ref{Fig2}, compared to classical communication systems, the semantic communication system introduces the semantic transmitter, which extracts the source semantic feature guided by the background knowledge and the specific task to be executed at the receiver. In this way, the meaning of the source information is formulated as semantic feature sequences in the semantic space. Due to this, as shown in Fig. \ref{Fig3}(b), each compressed sequence $X$ is associated with a typical ``fan'' of $S$ sequences that are jointly typical with the given $X$. This means all the source sequences in this fan corresponds to the identical semantics. From the perspective of classical source coding, this merging process of source typical sequences leads to lossy compression. However, due to the introduction of background knowledge, this process is indeed meaning-lossless since the semantics can be perfectly transmitted, or the task can be fully executed. Therefore, the number of sequences in the compact semantic space is obviously less than that in the classical compressed signal space.

On the other hand, for the channel transmission in semantic communications, as shown in Fig. \ref{Fig3}(b), since the feature sequences $X$ in the semantic space are of different importance, given the background knowledge in the receiver, the associated typical ``fan'' of each $X$ is of diverse sizes. For a more important semantic feature $X$, its typical ``fan'' of $Y$ sequences is of smaller size, i.e., its uncertainty on the received sequences is relatively small. However, for a less important semantic feature $X$, its typical ``fan'' of $Y$ sequences is of larger size, which corresponds to larger uncertainty. Therefore, the classic definition of channel capacity should be extended in semantic transmission.

In a nutshell, due to the introduction of semantic space, semantic communications can expand the theoretical scope of classical communications. The detailed framework of theoretical analysis should be further investigated.

\section{The Rate of Semantic Compression}

Shannon entropy based on probabilistic modeling is not appropriate to quantify semantic information. Alternatively, based on the computer framework, the semantics of messages turns out the ways of description by computer programs, skipping psychological and uncertain aspects. Computer programs process messages using discrete methods. In this section, we provide a brand new perspective of semantic compression stemming from the conditional descriptive complexity.

\subsection{Semantic Compression}

Alan M. Turing has given a brilliant demonstration that everything can be reasonably computed by a universal Turing machine, the primitive model of modern computers. Naturally, semantic information could also be modeled by a universal Turing machine. For example, the non-terminating binary sequence $11.001001000011...$ seems a random and meaningless sequence. However, it indeed stands for the decimal format of mathematical constant $\pi$. The descriptive complexity is not as high as it is presented technically. We can utilize Gauss-Legendre algorithm, which is simple and of finite-length, to print the infinite sequence. Hence, in semantic compression, abstract semantics is transformed into metrical information measured by the complexity of describing the message using a universal Turing Machine.

Based on the above assumptions, the rate of semantic compression can be quantified by algorithmic or descriptive complexity. Kolmogorov complexity \cite{li2008introduction}, a well-known metric in algorithmic information theory, is defined as the length of the shortest computer program used to print or describe the sequence. However, it is intractable to find the ideal shortest program in the global context in order to get the minimum rate of semantic compression. Hence, in practice, it is more rational to define the ``optimality'' over a local limited context. Huffman compressor captures the global property and builds codebook from training data. The same idea goes for semantic compression. Given the \emph{a priori} knowledge set, the semantic compression is viewed as a process of finding a shorter program conditioned on it to represent the message meaning. If the transceiver in semantic communications shares the same knowledge set, the coded sequence could be semantic-unambiguously decoded by transmitting the confusing parts up to an additive constant. The lowercase Greek letter $\pi$, the designate of the constant $3.1415...$, is transmitted without misunderstanding at the receiver as long as the equivalence is established between the two representations. As for this example, the number of symbols used to transmit the constant is yet apparently reduced. However, if the knowledge sets mismatch between the transceiver, the semantic transmission using current algorithm fails. Consequently, more symbols are required to clarify the message so that the receiver manages to get it across. Modern advances in AI enable powerful signal processing capability to formulate a shared background knowledge, which can alleviate such a mismatch problem.

\subsection{Semantic Compression Rate Bound}\label{section_limit}

As discussed before, if we view the semantic compression as a process of describing the message with a Turing machine, the semantic compression bound can be transformed to a kind of conditional algorithmic complexity. That is to say, the complexity of a message $x$ is the length of the shortest binary program to print it, given a specific knowledge base. Actually, due to the discrepancy between the message and the \emph{a priori} knowledge, the semantic compression is to eliminate the semantic uncertainty for reliable semantic communication. 


On that basis, we propose the concept \emph{normalized conditional complexity (NCC) bound} to measure the descriptive complexity of a sequence set with prior knowledge. The NCC bound is defined over a specific sequence set $\mathcal{D}_T$ conditioned on a knowledge base $\mathcal{D}$. Define $C$ as a function to compute descriptive complexity, then the NCC bound is a mathematical expectation and formulated as

\begin{equation} \label{EqNCC}
    \mathrm{NCC}\left(\mathcal{D}_T | \mathcal{D} \right)=  \mathop \mathbb{E}\limits_{x\in \mathcal{D}_T} {\left[\frac{C\left(x,\mathcal{D} \right)-C\left( \mathcal{D} \right)}{l(x)}\right]},
\end{equation}
where $C\left(x,\mathcal{D} \right)$ denotes the joint complexity. The numerator measures the extra complexity required to print the sequence $x$ with the aid of prior knowledge base, which is evidently not larger than its unconditional complexity $C\left(x \right)$. Then the conditional complexity is normalized by the sequence length $l\left (x \right)$. Finally, it takes expectation over all the sequences $x$ in $\mathcal{D}_T$. Given a collection of source messages, the NCC bound can be used to evaluate the descriptive complexity of the test set $\mathcal{D}_T$ conditioned on the training set $\mathcal{D}$. 

To compute unconditional descriptive complexity, the Kolmogorov complexity of a message essentially refers to the minimum length of the program to describe it, as stated before. However, the Kolmogorov complexity is non-computable \cite{li2008introduction}. It is intractable to find such an optimal program to describe a message in either conditional or unconditional cases. Therefore, in practice, a general compressor will be utilized as an approximation for practical application.

\begin{figure}[t]
  \centering
  \includegraphics[scale=0.6]{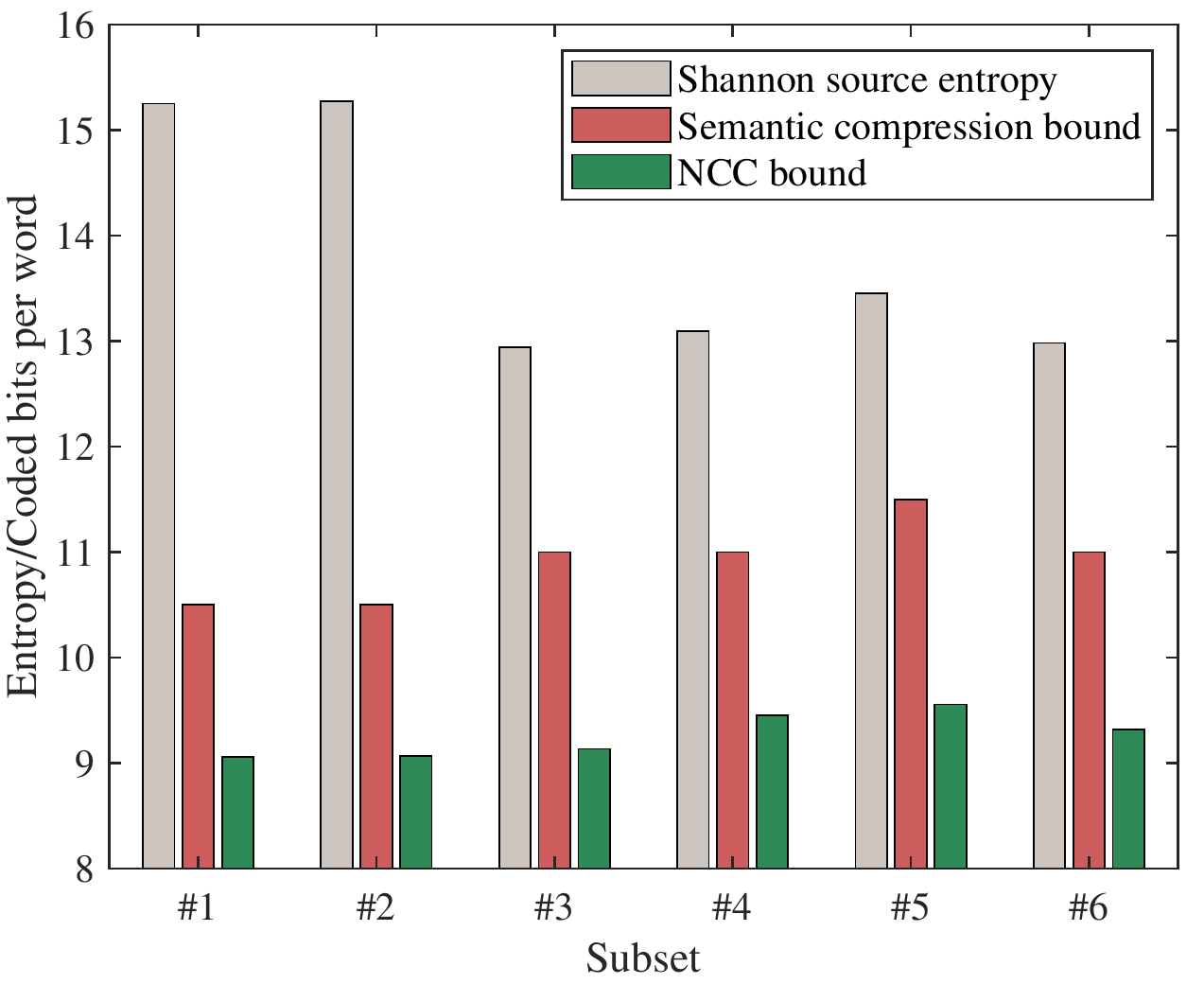}
  \caption{Evaluation of semantic compression rate on several corpora.}
  \label{Fig4}
\end{figure}

Evaluation of NCC bound on various text corpora in English is shown in Fig. \ref{Fig4}. The first two subsets are sampled from the press releases and legislative documents. The last ones are sampled from novel books of different sub-genres. The size of each subset is around millions of characters. In addition, the Shannon source entropy is provided, which is modeled as a first-order Markov source. Initial results for semantic compression of text are also presented in this figure, situating between the compression bound and the Shannon source entropy. The semantic compressor is based on Transformer neural network architecture \cite{vaswani2017attention}, and its detail will be given in the next section.

In a nutshell, the proposed NCC bound is a theoretical lower bound on the semantic compression rate, which does not rely on any semantic compression model. It is impossible to quantify how far off the computable NCC bound is from the intractable conditional Kolmogorov complexity. This NCC bound is a potential exploration on the semantic compression rate from the novel perspective of algorithmic information theory. Other perspectives and derivations of the semantic compression bound are still open to be explored.

\section{Exemplary Use Case of Semantic Communication}

Multiple sets of performance assessment metrics have been developed to guide the design of practical semantic communication systems, categorized by the modality of data source. For text files, the bilingual evaluation understudy (BLEU) \cite{deepsc}, word error rate (WER) \cite{farsad2018deep} are direct metrics to evaluate the accuracy of text transmission, and latent representation similarity derived from AI models is a novel metric to distinguish semantic similarity. For speech and audio sources \cite{speechqualityhandbook}, common metrics include objective metrics like perceptual evaluation of speech quality (PESQ), subjective scores like mean opinion score (MOS), and other metrics involved in downstream tasks. For image and video sources, human perception of image quality is taken into account \cite{ding2021comparison, rassool2017vmaf}, including peak-signal-to-noise ratio (PSNR), the multi-scale extension of the structural similarity index (MS-SSIM), the learned perceptual image patch similarity model (LPIPS), and the video multi-method assessment fusion (VMAF).

Following the above established performance assessment metrics, one can optimize the semantic communication system using the end-to-end transmission rate-distortion loss function. In this article, we demonstrate a use case of semantic communication for text transmission. Given the transmission rate, the semantic text transmission is intended to minimize the WER over the unpredictable wireless channel. 

\begin{figure}[t]
	\setlength{\abovecaptionskip}{0.cm}
	\setlength{\belowcaptionskip}{-0.cm}
	\centering{\includegraphics[scale=0.6]{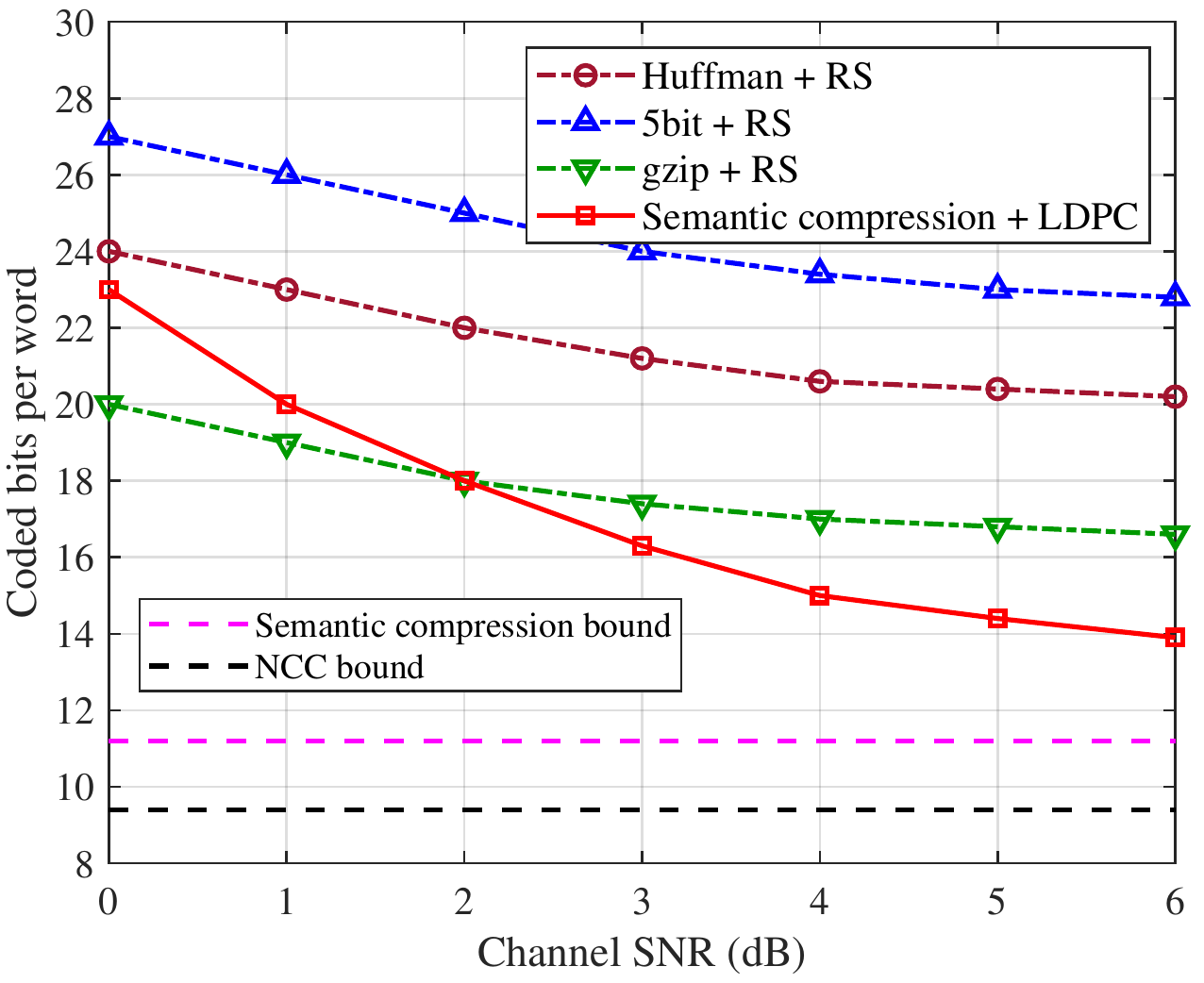}}
	\caption{Comparison of coded bits per word for reliable text transmission over the additive white Gaussian noise (AWGN) channels.}
    \label{Fig5}
\end{figure}

We note that aforementioned works explored the semantic communication system for text transmission using deep neural networks \cite{deepsc}. The issue is that every word shares the same semantic space with identical dimension. However, we note that these representation vectors embedded into the semantic space are redundant for some words. Our use case in this article sufficiently considers the contribution of words to the semantics of the whole phrase, sentence, and paragraph. A semantic compression model based on Transformer neural network \cite{vaswani2017attention} is designed to compress the text with around 11 bits per word, as shown by the dotted line ``Semantic compression bound" in Fig. \ref{Fig5}. Coded bits of each word are of different lengths, achieving variable-length semantic coding of text.

The performance of semantic coded transmission system is evaluated over the AWGN channel. Fig. \ref{Fig5} also compares the transmitted coded bits (source and channel coded bits) averaged per word for reliable transmission of source text. As a comparison, the baseline schemes of source coding include the fixed 5bit encoding for characters, Huffman coding, and gzip coding (dashed lines). The text decoding fails if the channel coding does not correct perfectly, in the case of gzip or Huffman codes. Following \cite{farsad2018deep}, Reed-Solomon (RS) channel coding is adopted, and the number of parity bits is optimized to fit the bit budget.

The semantic compression bound is slightly higher than the theoretical NCC bound since the semantic compression bound is tied with the selected neural network model. A smaller gap indicates a better performance of the semantic compression algorithm. With the increase of channel SNR, the number of parity-check bits required in channel coding decreases. As a result, the average number of source-channel coded bits to ensure reliable transmission also decreases. The solid line ``Semantic compression + LDPC'' in Fig. \ref{Fig5} shows the performance of our semantic compression coding combined with LDPC coding. It quickly approaches the semantic compression bound, which stands for the minimum number of coded bits required using the same semantic compression model. The minimum coded length to ensure reliable transmission in classical coding schemes is evidently higher than the NCC bound, as well as the compression bound constrained by the given semantic compression model.

Although finding out a better semantic compression model to approach the NCC bound still requires further works, these initial results verify that our semantic scheme can take much less channel bandwidth to correctly transmit the source text.

\section{Conclusion and Forward Looking}

This article has presented a brief tutorial on the framework of semantic communications, the gains analyzed from information theory and an initial exploration of the semantic compression bound. 

To pave the way to semantic communications, some open issues should be further addressed. On the one hand, beyond the constraint of semantic compression models, it is challenging to derive a universal theory to well bound the errors in source semantic compression and semantic-aware channel transmission. Furthermore, on the scene of large-scale communication networks, the theoretical analysis of multi-node semantic transmission or semantic networking deserves to be investigated. On the other hand, advanced performance evaluation models are expected to be extended for different machine-type purposes or tasks. In addition, the design criteria for multi-modal semantic communications will become gradually matured.

\ifCLASSOPTIONcaptionsoff
  \newpage
\fi

\bibliography{bliography}
\bibliographystyle{ieeetr}

\end{document}